\documentclass[conference]{IEEEtran}
\IEEEoverridecommandlockouts
\usepackage{cite}
\usepackage{amsmath,amssymb,amsfonts}
\usepackage{algorithmic}
\usepackage{graphicx}
\usepackage{textcomp}
\usepackage{xcolor}
\usepackage{multirow}
\usepackage{url}
\usepackage{tikz}
\usetikzlibrary{shapes.geometric, arrows.meta, positioning}
\def\BibTeX{{\rm B\kern-.05em{\sc i\kern-.025em b}\kern-.08em
    T\kern-.1667em\lower.7ex\hbox{E}\kern-.125emX}}
\begin{document}

\title{LLM-Based Robustness Testing of Microservice Applications:
An Empirical Study\\
{\large Practical Experience Report}}

\author{
\IEEEauthorblockN{Hrushitha Goud Tigulla}
\IEEEauthorblockA{College of Computing and Informatics\\
University of North Carolina at Charlotte\\
Charlotte, USA\\
htigulla@charlotte.edu}
\and
\IEEEauthorblockN{Marco Vieira}
\IEEEauthorblockA{College of Computing and Informatics\\
University of North Carolina at Charlotte\\
Charlotte, USA\\
marco.vieira@charlotte.edu}
}

\maketitle

\begin{center}
\small\textit{This work has been submitted to the IEEE for possible 
publication. Copyright may be transferred without notice, 
after which this version may no longer be accessible.}
\end{center}

\begin{abstract}
Malformed, missing, or boundary-value inputs in microservice APIs can cascade across dependent services, threatening reliability. Robustness testing systematically exercises such inputs to expose server-side failures, but generating diverse, effective tests remains challenging. Large Language Models can generate such tests from API specifications; however, it is unknown whether different models and prompt strategies produce diverse failure sets or converge on the same failures. {We report a controlled experiment applying} 7 prompt strategies to 3 open-source LLMs (14B--70B parameters) targeting 2 architecturally distinct microservice systems{: one Java monolingual (6~services, 9~failure modes) and one polyglot (27~services, 14~failure modes), yielding} 38 valid runs and 663 generated tests. {We find that} prompt strategy explains more variation in diversity than model size{: a Structured prompt collapses diversity entirely, while a single model varied across three prompt strategies achieves complete failure-mode coverage on one system, outperforming any multi-model ensemble under a fixed prompt.} We introduce two strategies, Guided and GuidedFewShot{,} that embed {a} mutation taxonomy {from prior robustness testing research} as domain context. GuidedFewShot achieves the highest single-run coverage on both systems (5 of 9 and 8 of 14 failure modes) while maintaining low cross-model similarity. {A key lesson is that taxonomy rules alone are insufficient: LLMs cannot distinguish key-absent from value-empty mutations without concrete examples.} Findings replicate across both systems.
\end{abstract}


\section{Introduction}
\label{sec:intro}

{Microservice architectures decompose applications into independently deployable services that communicate over HTTP and messaging protocols, each exposing endpoints that must handle not only well-formed requests but also malformed, missing, and boundary-value inputs without producing server-side failures.} Such failures compromise service reliability: a single unhandled input in one microservice can cascade through dependent services, degrading or disrupting the entire application. Robustness testing, the systematic exercise of such abnormal inputs, is well established for distributed systems. Mutation-based approaches provide structured taxonomies of input faults~\cite{vieira2007}, and tools like Ballista~\cite{kropp1998} demonstrated that automated fault injection can expose failures that conventional testing misses. More recently, Large Language Models (LLMs) have been applied to generate test cases from API specifications {~\cite{kim2024,testforge2025}}, raising a practical question: \textit{when different models and prompt strategies generate robustness tests for the same system, do they expose different failure modes, or do they all find the same bugs?}

The answer matters for testing practice. If every model-prompt configuration converges on the same failures, running multiple configurations wastes effort. If they diverge, practitioners can combine them to cover more failure modes at modest cost. Recent work has begun to systematically evaluate LLM-generated tests. For example, TestForge~\cite{testforge2025} introduced a benchmarking framework that measures the success of compilation, functional correctness, and code coverage of LLM-generated JUnit tests across prompting strategies and models. That work confirms LLMs can produce functional tests for well-defined problems but finds they struggle with behavioral diversity. {Existing studies of LLM-based test generation focus on code correctness, compilation rate, and line coverage~\cite{testforge2025,fraser2011}. None measure the diversity of failure modes discovered across models and prompt strategies in robustness testing, and the relationship between prompt design and failure-mode diversity has not been studied.}

Our study addresses a complementary question: rather than measuring whether generated tests are correct, we measure whether different configurations \textit{discover different failures} in running microservice systems. {We report a controlled empirical study based on practical experience applying LLM-based test generation to two open-source microservice systems.} We apply 7 prompt strategies to 3 open-source LLMs (14B, 32B, and 70B parameters, including one code-specialized model) targeting 2 architecturally distinct microservice systems: a Java monolingual application with 6 services and a polyglot application with 27 services written in Go, C++, Rust, JavaScript, .NET, and other languages. The experiment yields 38 valid runs and 663 generated tests. We define failure modes (FMs) as unique (endpoint, parameter, mutation type) triples verified against the application source code, identifying 9 FMs in the first system and 14 in the second. Coverage is measured as the fraction of FMs triggered per run. Diversity is measured as pairwise Jaccard similarity of FM sets across models using the same prompt strategy. We also investigate whether injecting domain knowledge from an established mutation taxonomy~\cite{vieira2007} into LLM prompts changes coverage and diversity.

The study addresses three research questions:

\noindent\textbf{RQ1 (Coverage):} How does failure mode coverage vary across models and prompt strategies?

\noindent\textbf{RQ2 (Diversity):} How diverse are the failure mode sets produced by different models under the same prompt strategy?

\noindent\textbf{RQ3 (Domain Knowledge):} Does incorporating mutation taxonomy rules from robustness testing research into LLM prompts improve coverage without collapsing diversity?

\smallskip
The main contributions are:

\begin{itemize}
    \item {Empirical evidence that the prompt strategy explains more diversity variation than model size. Changing the prompt for a fixed model produces greater FM set variation than changing the model under a fixed prompt.}

    \item {Identification and cross-system replication of \textit{Structured collapse}, where a prescriptive prompt causes all three models to produce identical FM sets (Jaccard $= 1.00$ on both systems).}

    \item {Design of two new prompt strategies, Guided and GuidedFewShot, that embed the mutation taxonomy from Vieira et al.~\cite{vieira2007} directly into the LLM prompt as domain context, providing mutation rules without prescribing specific test inputs.}

    \item {Evidence that domain knowledge transforms a zero-coverage model-strategy pair into the highest-performing configuration. A code-specialized model that triggered zero robustness failures under Self-Refine achieved 5 of 9 FMs (56\%) and 8 of 14 FMs (57\%) under GuidedFewShot, the highest single-run coverage on both systems.}

    \item {Evidence that prompt variation for a single model outperforms model variation under a fixed prompt: one model varied across three strategies achieves complete FM coverage on one system, while the best single-strategy multi-model ensemble reaches only 64\%.}

    \item {Cross-system replication of all findings across two architecturally different microservice applications, covering monolingual versus polyglot implementations and form-encoded versus JSON communication protocols.}
\end{itemize}

This paper is organized as follows. Section~\ref{sec:related} reviews related work. Section~\ref{sec:design} describes the study design. Section~\ref{sec:fm} defines failure modes. Section~\ref{sec:results} presents results by research question. Section~\ref{sec:discussion} discusses implications. Section~\ref{sec:threats} addresses threats to validity, and Section~\ref{sec:conclusion} concludes. 

\section{Related Work}
\label{sec:related}

Automated robustness testing originated with Ballista~\cite{kropp1998}, which tested POSIX interfaces by substituting invalid values drawn from type-specific fault models. The same work introduced the CRASH failure classification (Catastrophic, Restart, Abort, Silent, Hindering) as a severity scale for observed failures. Vieira et al.~\cite{vieira2007} extended this approach to web services, organizing mutations by data type: string (null, empty, nonprintable, overflow), numeric (null, zero, negative, boundary), list (null, empty, element removal, duplication), and structural (field-absent, field-null, empty body). These ideas were later operationalized for SOAP services by wsrbench~\cite{laranjeiro2008} and for REST services by black-box testing tools~\cite{laranjeiro2021}. {Our Guided prompt strategies embed this mutation taxonomy directly as domain context for LLM-based test generation, extending its applicability to a new test generation paradigm.}

Stateful REST API testing has been addressed by RESTler~\cite{atlidakis2019}, which infers producer-consumer dependencies between API operations and uses them to generate request sequences that exercise deeper application logic. EvoMaster~\cite{arcuri2019} applies evolutionary search to REST API testing, optimizing for code coverage and fault detection. These tools generate tests programmatically from API schemas. Zhang and Arcuri~\cite{zhang2023} compared seven fuzzers on 18 APIs, finding that existing tools detect only a small subset of possible fault types, with most relying on HTTP 500 status codes as the primary oracle. Our work uses the same oracle but studies how \textit{LLM-based} generation compares across model-prompt configurations rather than across dedicated fuzzing tools.

Recent studies have applied large language models to unit test generation~\cite{kim2024,testforge2025}, finding that LLMs produce compilable tests at high rates and can achieve line coverage comparable to search-based tools such as EvoSuite~\cite{fraser2011} and Randoop~\cite{pacheco2007} on some benchmarks. TestForge~\cite{testforge2025} introduced a benchmarking framework that evaluates compilation success, runtime behavior, semantic validity, and functional correctness of LLM-generated JUnit tests. Across four open-source models and three prompting strategies, the study found that code-specialized models produce structurally reliable but conservative test oracles, while larger general-purpose models generate more expressive assertions once compilation succeeds. These findings address test \textit{quality}. Our study addresses a different dimension: test \textit{diversity}, specifically, whether different model-prompt configurations reveal distinct robustness failures in running systems. Kim et al.~\cite{kim2024} used LLMs to improve REST API testing by generating test inputs from natural-language API descriptions, demonstrating improvements over schema-based fuzzing. These studies typically evaluate one or two models on a single system. {None measure pairwise diversity of failure-mode sets across configurations or investigate the relationship between prompt design and failure-mode diversity in robustness testing of microservice systems.}

Prompt engineering techniques have been studied across several code generation tasks. Zero-shot prompting provides only a task description. Few-shot prompting supplies worked examples that anchor model output~\cite{brown2020}, and comparative studies generally find that it improves output quality. Chain-of-thought (CoT) prompting~\cite{wei2022} elicits step-by-step reasoning before the final output, which helps with multi-step logic. Self-refinement strategies take a different approach: the model critiques and revises its own generations~\cite{madaan2023}. These comparisons measure correctness and syntactic quality of generated code, not diversity of test outcomes. {Our study evaluates all of these strategies, plus two new domain-knowledge strategies grounded in the mutation taxonomy of Vieira et al.~\cite{vieira2007}, and measures their effect on robustness failure-mode diversity rather than code quality alone. To the best of our knowledge, no prior work has examined the combined effect of prompt strategy and model choice on the diversity of robustness failures observed when running microservice systems.}

\section{Study Design}
\label{sec:design}

{This section describes the experimental setup used in our study.} Table~\ref{tab:config} summarizes the experiment matrix.

\subsection{Systems Under Test}
\label{sec:suts}

We use two microservice applications with different architectures, languages, and protocols. The differences support cross-system replication of findings.

TeaStore~\cite{vonkistowski2018} is a Java-based microservice reference application with 6 services deployed as Docker containers. It implements an online tea shop with product browsing, shopping cart management, user authentication, and a recommendation engine. Services communicate via form-encoded HTTP POST requests. The application exposes endpoints for retrieving products, adding cart items, and removing cart items.

The OTel Astronomy Shop~\cite{otel2024} is the OpenTelemetry community demonstration application, a polyglot system with 27 services written in Go, C++, Rust, JavaScript, .NET, Java, Python, and PHP. It implements an online astronomy equipment store with a product catalog, a shopping cart, and a multi-field checkout. Services communicate via JSON APIs and internal gRPC calls.

The two systems differ on every architectural dimension relevant to robustness testing: language homogeneity (monolingual versus polyglot), service count (6 versus 27), request encoding (form-encoded versus JSON), and input complexity (single-parameter operations versus a checkout endpoint with credit card, address, and currency fields). {Both are widely used open-source benchmarks with fully available source code, enabling the empirical FM verification described in Section~\ref{sec:fm}. Their architectural diversity, rather than industrial provenance, supports cross-system replication of findings.}

\begin{table}[t]
\centering
\caption{Experimental Configuration}
\label{tab:config}
\renewcommand{\arraystretch}{1.2}
\begin{tabular}{p{1.1cm} p{2.6cm} p{3.9cm}}
\hline
\textbf{Component} & \textbf{Element} & \textbf{Details} \\
\hline
\multirow{2}{*}{SUTs}
  & TeaStore & Java, 6 services, form-encoded HTTP, 9~FMs \\
  & OTel Demo & Polyglot, 27 services, JSON APIs, 14~FMs \\
\hline
\multirow{3}{*}{Models}
  & {\texttt{qwen3:14b}} & 14B params, general-purpose \\
  & {\texttt{qwen2.5-coder:32b}} & 32B params, code-specialized \\
  & {\texttt{llama3.1:70b}} & 70B params, general-purpose \\
\hline
Oracle & HTTP status & Response code $\geq$ 500 \\
\hline
Runs & 38 valid & of 42 possible (7 $\times$ 3 $\times$ 2) \\
\hline
\end{tabular}
\vspace{-12pt}
\end{table}

\subsection{Models}
\label{sec:models}

We evaluate three open-source LLMs hosted locally via Ollama with default inference parameters. Running models locally ensures reproducibility without reliance on commercial APIs, rate limits, or costs. {\texttt{qwen3:14b}} is a 14B-parameter general-purpose model. {\texttt{qwen2.5-coder:32b}} is a 32B-parameter code-specialized model trained with emphasis on code generation. {\texttt{llama3.1:70b}} is a 70B-parameter general-purpose model, the largest in our study. The three models span a 5$\times$ parameter range and include one code-specialized model, allowing us to assess the effects of model size and code-specific training on robustness test diversity.

{Each model-prompt-SUT configuration is executed once. We acknowledge this limitation and discuss its implications in Section~\ref{sec:threats}.}

\subsection{Prompt Strategies}
\label{sec:strategies}

We evaluate 7 prompt strategies in three groups. Table~\ref{tab:strategies} describes each one.

The first group provides two baselines. ZeroShot provides only the API specification, with minimal instructions for generating robustness tests. Structured provides a prebuilt violation table that maps each endpoint to specific mutation types and expected outcomes, effectively a test specification that the model must translate into code. {This table was constructed manually by the authors for each SUT, enumerating endpoints, applicable mutation types from the Vieira et al.\ taxonomy, and expected HTTP outcomes.}

The second group applies three established prompting techniques. FewShot supplies 3--4 working test examples demonstrating robustness failure patterns {, then instructs the model to generate additional tests following the same patterns}~\cite{brown2020}. Chain-of-Thought (CoT) instructs the model to reason step-by-step through violation categories before producing test code~\cite{wei2022}. Self-Refine implements a three-step cycle: generate a draft, critique it, then produce a revised version~\cite{madaan2023}.

The third group introduces two new strategies built on domain knowledge from robustness testing research. Guided includes the mutation taxonomy from Vieira et al.~\cite{vieira2007}, organized by data type (string, numeric, list, structural). It provides mutation rules as context but includes no examples and does not constrain the number of tests. GuidedFewShot extends Guided with the CRASH failure classification and concrete examples showing the distinction between key-absent mutations (the parameter key is not present in the HTTP body), value-empty mutations (the key is present, but the value is an empty string), and value-null mutations (the key is present with a null value). This distinction is critical because {preliminary results showed} all three models, given only the taxonomy rules, interpreted ``replace by null'' as setting a value to the empty string rather than removing the key entirely.

{The design rationale for these prompting strategies addresses a tension observed in practice.} Structured prompting achieves consistent coverage but kills diversity: all models produce identical FM sets. The mutation taxonomy provides domain knowledge at a higher level of abstraction, without prescribing specific inputs. Guided tests whether rules alone help. GuidedFewShot tests whether concrete examples resolve the ambiguity that the rules leave open.

\begin{table}[t]
\centering
\caption{Prompt Strategies}
\label{tab:strategies}
\renewcommand{\arraystretch}{1.2}
\begin{tabular}{p{1.8cm} p{1.0cm} p{4.0cm}}
\hline
\textbf{Strategy} & \textbf{Group} & \textbf{Description} \\
\hline
ZeroShot & Base & API specification only; no examples, rules, or guidance \\
Structured & Base & Pre-built violation table mapping endpoints to mutations and expected outcomes \\
\hline
FewShot & Estab. & 3--4 working test examples demonstrating failure patterns \\
CoT & Estab. & Step-by-step reasoning through violation categories before code generation \\
Self-Refine & Estab. & Three-step draft $\rightarrow$ self-critique $\rightarrow$ rewrite cycle \\
\hline
Guided & Domain & Mutation taxonomy rules by data type; no examples, no test count constraint \\
GuidedFewShot & Domain & Mutation taxonomy + CRASH classification + examples of key-absent vs.\ value-empty \\
\hline
\end{tabular}
\vspace{-12pt}
\end{table}

\subsection{Experimental Pipeline}
\label{sec:pipeline}

\begin{figure}[b]
\vspace{-12pt}
\centering
{%
\begin{tikzpicture}[
  node distance=0.45cm and 0.1cm,
  box/.style={rectangle, rounded corners=3pt, draw=black, fill=gray!10,
              text width=2.5cm, align=center, font=\small, minimum height=0.7cm},
  arr/.style={-{Stealth[length=4pt]}, thick}
]
\node[box] (step1) {1. Prompt\\ Construction};
\node[box, below=of step1] (step2) {2. LLM\\ Generation};
\node[box, below=of step2] (step3) {3. Output\\ Cleaning};
\node[box, below=of step3] (step4) {4. Test\\ Execution};
\node[box, below=of step4] (step5) {5. Response\\ Classification};
\draw[arr] (step1) -- (step2);
\draw[arr] (step2) -- (step3);
\draw[arr] (step3) -- (step4);
\draw[arr] (step4) -- (step5);
\node[right=0.5cm of step1, font=\scriptsize, text width=3.6cm]
  {API spec + strategy rules/examples};
\node[right=0.5cm of step2, font=\scriptsize, text width=3.6cm]
  {Java/JUnit 5 test code};
\node[right=0.5cm of step3, font=\scriptsize, text width=3.6cm]
  {Remove markdown, strip reasoning blocks; exclude if compile fails};
\node[right=0.5cm of step4, font=\scriptsize, text width=3.6cm]
  {Live SUT in Docker};
\node[right=0.5cm of step5, font=\scriptsize, text width=3.6cm]
  {HTTP $\geq$ 500 $\Rightarrow$ robustness failure (Abort class)};
\end{tikzpicture}%
}
\caption{{Experimental pipeline.}}
\label{fig:pipeline}
\end{figure}
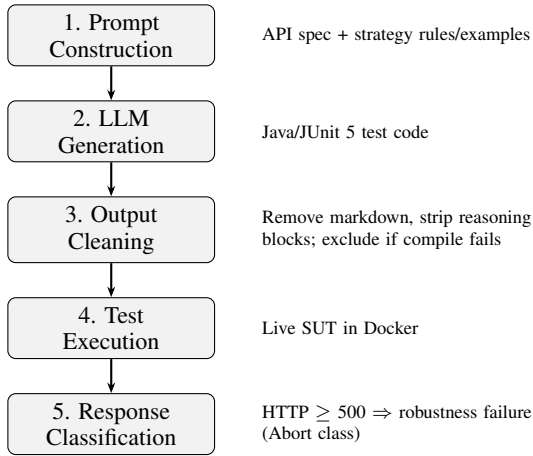

Each run follows five steps, as illustrated in Fig.~\ref{fig:pipeline}. First, the prompt for the selected strategy is constructed with the SUT's API specification and, where applicable, mutation taxonomy rules and examples. Second, the prompt is submitted to the model, which generates {Java test code using JUnit 5 and the \texttt{java.net.http} HTTP client}. Third, the output is cleaned: markdown formatting artifacts are removed, and reasoning blocks that some models produced despite explicit instructions to output only test code are stripped. Runs that fail to compile after cleaning are excluded. Fourth, compiled tests are executed against a live instance of the SUT deployed in Docker. Fifth, HTTP responses are classified: any response with a status code $\geq$ 500 is recorded as a robustness failure, corresponding to the Abort class in the CRASH taxonomy~\cite{vieira2007}.

Of the 42 possible configurations (7 strategies $\times$ 3 models $\times$ 2 SUTs), 38 produced valid runs. Four were excluded: {\texttt{qwen3:14b}} under ZeroShot on both SUTs (the model switched to a non-English language mid-generation, producing unusable output), {\texttt{qwen3:14b}} under Self-Refine on OTel (malformed output that could not be parsed), and {\texttt{llama3.1:70b}} under Structured on OTel (repeated system crashes during generation). {Excluded runs were not retried, consistent with the single-execution design. The language switch is a known instability in multilingual models when no language is specified; the generation crashes reflect memory exhaustion during long-context generation for Structured prompts on the larger system.} 

{Failure mode definitions were derived empirically. After executing all runs, we identified distinct (endpoint, parameter, mutation-type) triples that triggered robustness failures and verified each against the application source code, yielding 9 FMs for TeaStore and 14 for OTel. We discuss the implications of this post-hoc derivation in Section~\ref{sec:threats}. Section~\ref{sec:fm} defines the FMs in detail.} \textcolor{black}{We focus on the Abort class because it is the only CRASH class observable through HTTP status codes alone; detecting other classes such as Complete or Hindering would require application-level monitoring beyond this study.}

\subsection{Diversity Metrics}
\label{sec:metrics}

We measure diversity using Jaccard similarity on binary FM coverage vectors. For each run, we construct a binary vector of length equal to the number of FMs for that SUT, where entry $i$ is 1 if the run triggered FM$_i$ and 0 otherwise. The Jaccard similarity between two runs $A$ and $B$ is:
\begin{equation}
J(A, B) = \frac{|A \cap B|}{|A \cup B|}
\label{eq:jaccard}
\end{equation}
\noindent where $A$ and $B$ are the sets of triggered FMs. A value of 1.00 means identical FM sets; 0.00 means completely disjoint sets. We report the average pairwise Jaccard index across all model pairs within each prompt strategy to quantify the diversity a strategy preserves. Note that, {with only three models per strategy, each strategy yields at most three pairwise Jaccard values. We therefore treat these values as descriptive indicators rather than statistically significant estimates, and discuss this limitation further in Section~\ref{sec:threats}.}

We also compute union coverage: the set union of FMs triggered across all runs of a given strategy on a given SUT. Union coverage captures the total FM space reachable when multiple models are available. The gap between the best single-run coverage and the union coverage indicates the extent of the benefit a multi-model ensemble provides.

\section{Failure Mode Definitions}
\label{sec:fm}

A failure mode (FM) is a unique (endpoint, parameter, mutation-type) triple that triggers an HTTP $\geq$ 500 response. Each FM maps to a mutation from the taxonomy of Vieira et al.~\cite{vieira2007} applied to a specific parameter of a specific endpoint. We define the 9 TeaStore FMs (Table~\ref{tab:ts_fm}) and the 14 OTel FMs (Table~\ref{tab:otel_fm}). {The definitions were derived empirically:} {after executing all 38 valid runs, we collected every distinct input mutation that triggered a robustness failure and mapped it to an (endpoint, parameter, mutation-type) triple. Each triple was verified against the application source code to confirm that the mutation caused the 500-level response. We discuss the implications of this post-hoc derivation in Section~\ref{sec:threats}.}\footnote{The complete binary coverage matrices, test source files, and execution results for all 38 runs are available in an anonymous repository: \url{https://anonymous.4open.science/r/SRDS2026-replications-26B2/README.md}}

\subsection{TeaStore Failure Modes}
\label{sec:ts_fm}

Table~\ref{tab:ts_fm} lists the 9 TeaStore FMs. The product endpoint accepts an \texttt{id} parameter; supplying a non-numeric string (FM$_1$) or an empty string (FM$_2$) triggers a server-side parsing exception. The cart addition and removal endpoints accept a \texttt{productid} parameter in the form-encoded HTTP body.

A distinction critical to the Guided prompting findings is the difference between key-absent and value-empty mutations. FM$_3$ and FM$_6$ are key-absent: the \texttt{productid} key does not appear in the request body at all (e.g., \texttt{body: addToCart=\&} with no \texttt{productid} field). FM$_4$ and FM$_7$ are value-empty: the key is present but set to an empty string (e.g., \texttt{body: addToCart=\&productid=}). Java's \texttt{request.getParameter()} returns \texttt{null} for a missing key and an empty string for an empty value. Both produce 500-level responses, but through different exception chains. {This matters because LLMs, given only the taxonomy rule ``replace parameter value by null,'' consistently generate value-empty mutations and never generate key-absent mutations unless shown concrete examples, as we discuss in Section~\ref{sec:results}.}

FM$_9$ is the only authentication-related FM: submitting a valid product ID to the cart removal endpoint without an authenticated session triggers a \texttt{NullPointerException} on the session object. {This FM is found by only 3 of 20 TeaStore runs, exclusively by strategies that do not prescribe specific parameter mutations, a pattern we revisit in Section~\ref{sec:results}.}

\begin{table}[t]
\centering
\caption{TeaStore Failure Modes (FM$_1$--FM$_9$)}
\label{tab:ts_fm}
\renewcommand{\arraystretch}{1.15}
\begin{tabular}{c l l l}
\hline
\textbf{FM} & \textbf{Endpoint} & \textbf{Parameter} & \textbf{Mutation} \\
\hline
1  & GET /products    & id        & Non-numeric string \\
2  & GET /products    & id        & Empty string \\
3  & POST /cartAdd    & productid & Key absent \\
4  & POST /cartAdd    & productid & Value empty \\
5  & POST /cartAdd    & productid & Invalid string \\
6  & POST /cartRemove & productid & Key absent \\
7  & POST /cartRemove & productid & Value empty \\
8  & POST /cartRemove & productid & Non-existent ID \\
9  & POST /cartRemove & session   & Unauthenticated \\
\hline
\end{tabular}
\vspace{-12pt}
\end{table}

\subsection{OTel Astronomy Shop Failure Modes}
\label{sec:otel_fm}

Table~\ref{tab:otel_fm} lists the 14 OTel FMs. The product endpoint accepts a path parameter \texttt{id}; an invalid format string (FM$_1$) or an extremely long string of 500+ characters (FM$_2$) each triggers a failure. The cart endpoint accepts a JSON body with an \texttt{item} field; null (FM$_3$) and missing (FM$_4$) are distinct mutations exercising different server-side validation paths. The checkout endpoint accepts a JSON body with 12 fields. FM$_5$ through FM$_{14}$ each mutate one checkout field.

An infrastructure limitation affects verification of the checkout FMs. The OTel currency service (POST /api/currency) and shipping service (POST /api/shipping) return HTTP 504 for all inputs in the available Docker deployment, regardless of validity. Because checkout internally calls both services, HTTP 500 responses to checkout requests may reflect the 504 cascade rather than the input mutation. We attempted multiple deployment configurations, including a minimal Docker Compose setup and a custom configuration with Kafka added and observability components stripped. In all configurations, the currency and shipping containers started but timed out through the Envoy proxy {(the service mesh proxy used by the OTel demo for inter-service routing)} due to a gRPC routing issue.

We classify checkout FMs (FM$_5$ through FM$_{14}$) by their intended input mutation, verified from the checkout service's Java source code. One checkout test under {\texttt{qwen2.5-coder:32b}} with GuidedFewShot returned HTTP 200, confirming that the checkout service performs independent input validation before calling downstream services. This partial evidence supports the intention-based classification. {We acknowledge this as a limitation of the study and discuss its implications in Section~\ref{sec:threats}.}

\begin{table}[t]
\centering
\caption{OTel Astronomy Shop Failure Modes (FM$_1$--FM$_{14}$)}
\label{tab:otel_fm}
\renewcommand{\arraystretch}{1.15}
\begin{tabular}{c l l l}
\hline
\textbf{FM} & \textbf{Endpoint} & \textbf{Field} & \textbf{Mutation} \\
\hline
1  & GET /products/\{id\}  & id             & Invalid format \\
2  & GET /products/\{id\}  & id             & Very long (500+) \\
3  & POST /cart            & item           & Null value \\
4  & POST /cart            & item           & Field missing \\
5  & POST /checkout        & userId         & Empty string \\
6  & POST /checkout        & userCurrency   & Invalid code \\
7  & POST /checkout        & cc.number      & Empty string \\
8  & POST /checkout        & cc.year        & Expired value \\
9  & POST /checkout        & address.street & Empty string \\
10 & POST /checkout        & email          & Invalid format \\
11 & POST /checkout        & cc.number      & Wrong format \\
12 & POST /checkout        & userCurrency   & Empty string \\
13 & POST /checkout        & cc.month       & Negative value \\
14 & POST /checkout        & cart           & Empty/invalid \\
\hline
\end{tabular}
\vspace{-12pt}
\end{table}

{One run ({\texttt{llama3.1:70b}} under GuidedFewShot on OTel) generated a cart quantity mutation using \texttt{Integer.MAX\_VALUE + 1} {, which causes an integer overflow producing an unexpected negative value that the server fails to handle,} triggering a robustness failure outside FM$_1$ through FM$_{14}$. We do not add this as FM$_{15}$ to preserve denominator consistency across all 38 runs, since only the GuidedFewShot runs had the domain context needed to surface this mutation. This observation suggests that taxonomy-guided generation has exploratory value beyond reproducing known failures, and we return to it in Section~\ref{sec:discussion}.}

\section{Results}
\label{sec:results}

{This section presents results by research question. Given the small number of models per strategy (three), all Jaccard values should be interpreted as indicators rather than statistically significant estimates.} Table~\ref{tab:coverage} summarizes coverage by prompt strategy across both SUTs. {Table~\ref{tab:model_union} reports within-model cross-prompt union coverage.} Table~\ref{tab:jaccard} reports average Jaccard similarity per strategy. Table~\ref{tab:findings} lists all 10 findings.

\subsection{RQ1: Coverage}
\label{sec:rq1}

\subsubsection{Per-Strategy Coverage}

Table~\ref{tab:coverage} reports the best single-run coverage and union coverage for each prompt strategy. No single run achieves complete coverage on either system. The highest single-run coverage is 5 of 9 FMs (56\%) on TeaStore, achieved by {\texttt{qwen3:14b}} under FewShot and {\texttt{qwen2.5-coder:32b}} under GuidedFewShot. On OTel, the highest single-run coverage is 8 of 14 FMs (57\%), achieved by {\texttt{qwen2.5-coder:32b}} under GuidedFewShot. Across all 38 runs, no configuration exceeds 57\% {({Finding~1}; all findings are summarised in Table~\ref{tab:findings})}..

Union coverage varies widely. On TeaStore, FewShot achieves the highest union at 8 of 9 FMs (89\%): running all three models with FewShot prompts collectively triggers all but one FM. On OTel, GuidedFewShot achieves the highest union at 9 of 14 FMs (64\%). Structured reaches 7 of 14 (50\%) on OTel but only 2 of 9 (22\%) on TeaStore, indicating that prescribing violation types does not guarantee high coverage.

ZeroShot performs poorly overall: 1 of 9 FMs (11\%) on TeaStore and 5 of 14 (36\%) union on OTel. Without guidance, models generate syntactically valid inputs that do not exercise boundary conditions or structural mutations. CoT achieves 7 of 9 FMs (78\%) on TeaStore and 7 of 14 (50\%) on OTel, indicating that step-by-step reasoning helps models consider a broader range of mutations.

The gap between best single-run and union coverage reveals how much models complement each other under a given strategy. FewShot shows the largest gap on TeaStore: 56\% single-run versus 89\% union, a 33-point gain. Structured shows no gap (22\% single-run and 22\% union on TeaStore, 50\% and 50\% on OTel). This zero gap means all models produce identical outputs, a phenomenon examined in Section~\ref{sec:rq2}.

\subsubsection{ZeroShot Discovers State-Based FMs}

Despite low overall coverage, ZeroShot is the only strategy that discovers certain state-based FMs. On TeaStore, {\texttt{llama3.1:70b}} under ZeroShot is the only ZeroShot run to trigger any FM, finding FM$_9$ (unauthenticated cart removal), where the test removes a product without an authenticated session. FM$_9$ is found by only 3 of 20 TeaStore runs: {\texttt{llama3.1:70b}} ZeroShot, {\texttt{qwen3:14b}} CoT, and {\texttt{qwen3:14b}} Self-Refine. On OTel, the pattern repeats: both {\texttt{qwen2.5-coder:32b}} and {\texttt{llama3.1:70b}} under ZeroShot find FM$_{14}$ (checkout with empty/invalid cart), where the test attempts checkout before adding items. FM$_{14}$ appears in only 2 of 18 OTel runs, both ZeroShot. No guided, structured, or example-based strategy finds it.

The explanation is that parameter-focused strategies (Structured, Guided, GuidedFewShot) direct models toward field-level mutations: send an empty string here, remove a key there. ZeroShot models, with no such direction, generate tests that resemble realistic user journeys and naturally produce state-level mutations like ``check out without a cart'' or ``remove a product without logging in.'' {This suggests that even a carefully designed testing pipeline should include at least one ZeroShot run, since its low per-run coverage is offset by its ability to find FMs that are invisible to parameter-focused strategies.}

\subsubsection{Within-Model Cross-Prompt Union}
\label{sec:within_model}

The analysis above measures union coverage across models under a fixed strategy. {The complementary and more practical question is: how much coverage does a single model achieve when run with multiple prompt strategies? A practitioner is more likely to have one model with multiple prompts than multiple models with a single prompt.} Table~\ref{tab:model_union} reports the within-model cross-prompt union for each model. On TeaStore, {\texttt{qwen3:14b}} and {\texttt{llama3.1:70b}} each reach 8 of 9 FMs (89\%) across all their prompt runs, while {\texttt{qwen2.5-coder:32b}} reaches 7 of 9 (78\%). The OTel results are more striking: {\texttt{qwen2.5-coder:32b}} reaches 14 of 14 FMs (100\%), covering every known failure mode through prompt variation alone, compared to 11 of 14 (79\%) for {\texttt{qwen3:14b}} and 12 of 14 (86\%) for {\texttt{llama3.1:70b}}. 

{\texttt{qwen2.5-coder:32b}}'s 100\% coverage on OTel requires only three strategies: Structured (which contributes FM$_1$, FM$_2$, FM$_3$, FM$_5$, FM$_6$, FM$_7$, FM$_8$), GuidedFewShot (which adds FM$_9$, FM$_{10}$, FM$_{11}$, FM$_{12}$, FM$_{13}$), and ZeroShot (which adds FM$_4$ and FM$_{14}$, the state-based FM that only ZeroShot discovers). Three prompt strategies applied to a single model cover more FMs than any multi-model ensemble using a single strategy. {This result has a direct practical implication: a practitioner with access to only {\texttt{qwen2.5-coder:32b}} can achieve complete OTel FM coverage by varying the prompt, while running all three models with GuidedFewShot (the best single strategy) reaches only 64\%.} This strengthens {Finding~2}: prompt strategy matters more than model choice.

Each model has different blind spots. On TeaStore, {\texttt{qwen3:14b}} never finds FM$_3$ (key-absent on cartAdd), {\texttt{qwen2.5-coder:32b}} never finds FM$_5$ (invalid string productid) or FM$_9$ (unauthenticated session), and {\texttt{llama3.1:70b}} never finds FM$_6$ (key-absent on cartRemove). On OTel, FM$_{13}$ (negative card month) is the hardest FM: only {\texttt{qwen2.5-coder:32b}} triggers it, and only under FewShot and GuidedFewShot. No other model finds FM$_{13}$ under any strategy.

\begin{table}[t]
\centering
\caption{Coverage by Prompt Strategy}
\label{tab:coverage}
\renewcommand{\arraystretch}{1.15}
\begin{tabular}{l c c c c}
\hline
& \multicolumn{2}{c}{\textbf{TeaStore (9 FMs)}} & \multicolumn{2}{c}{\textbf{OTel (14 FMs)}} \\
\cline{2-3} \cline{4-5}
\textbf{Strategy} & \textbf{Best} & \textbf{Union} & \textbf{Best} & \textbf{Union} \\
\hline
ZeroShot       & 1 (11\%) & 1 (11\%) & 4 (29\%) & 5 (36\%) \\
Structured     & 2 (22\%) & 2 (22\%) & 7 (50\%) & 7 (50\%) \\
FewShot        & 5 (56\%) & 8 (89\%) & 4 (29\%) & 6 (43\%) \\
CoT            & 4 (44\%) & 7 (78\%) & 6 (43\%) & 7 (50\%) \\
Self-Refine    & 4 (44\%) & 5 (56\%) & 6 (43\%) & 6 (43\%) \\
Guided         & 2 (22\%) & 4 (44\%) & 6 (43\%) & 7 (50\%) \\
GuidedFewShot  & 5 (56\%) & 6 (67\%) & 8 (57\%) & 9 (64\%) \\
\hline
\end{tabular}
\vspace{-12pt}
\end{table}

\begin{table}[b]
\vspace{-12pt}
\centering
\caption{Within-Model Cross-Prompt Union Coverage}
\label{tab:model_union}
\renewcommand{\arraystretch}{1.15}

\begin{tabular}{p{2.6cm} p{1.2cm} p{1.6cm} p{1.8cm}}
\hline
\textbf{Model} & \textbf{TeaStore} & \textbf{OTel} & \textbf{Missing FMs} \\
\hline
{\texttt{qwen3:14b}} & 8/9 (89\%) & 11/14 (79\%) & TS: FM$_3$; OT: FM$_{10,13,14}$ \\
{\texttt{qwen2.5-coder:32b}} & 7/9 (78\%) & 14/14 (100\%) & TS: FM$_5$, FM$_9$ \\
{\texttt{llama3.1:70b}} & 8/9 (89\%) & 12/14 (86\%) & TS: FM$_6$; OT: FM$_{8,13}$ \\
\hline
\end{tabular}
\end{table}

\subsection{RQ2: Diversity}
\label{sec:rq2}

Table~\ref{tab:jaccard} reports the average pairwise Jaccard similarity across models for each strategy. Lower values mean greater diversity.
Structured prompting produces $J = 1.00$ on both SUTs: every model triggers exactly the same FMs ({Finding~3}). On TeaStore, all models trigger FM$_1$ and FM$_8$, nothing else. On OTel, all models trigger the same 7 FMs. The prompt leaves no room for interpretation, so model differences become irrelevant.

At the other extreme, Self-Refine on OTel produces $J = 0.00$: {\texttt{qwen3:14b}} and {\texttt{llama3.1:70b}} trigger completely disjoint FM sets, and {\texttt{qwen2.5-coder:32b}} {triggers zero FMs}. FewShot produces low Jaccard on both systems ($J = 0.16$ on TeaStore, $J = 0.31$ on OTel), indicating that worked examples anchor each model toward different failure regions.

The domain-knowledge strategies show different diversity profiles across SUTs. Guided achieves $J = 0.17$ on TeaStore and $J = 0.43$ on OTel. The higher OTel value reflects convergence toward the checkout endpoint's many fields. GuidedFewShot achieves $J = 0.38$ on TeaStore and $J = 0.12$ on OTel. The diversity effect is stronger on OTel, where GuidedFewShot is the most diverse strategy with $J > 0$: the combination of taxonomy rules and concrete examples causes each model to explore a different region of the FM space. On TeaStore, the three models share more FMs ($J = 0.38$), reflecting greater convergence on the smaller FM set.

A systematic comparison confirms {Finding~2} {(see Table~\ref{tab:findings})}. We computed average pairwise Jaccard similarity for all within-model cross-prompt pairs (same model, different prompts) and all cross-model same-prompt pairs (same prompt, different models). On TeaStore, the within-model cross-prompt average is $J = 0.16$, compared to $J = 0.34$ for cross-model same-prompt pairs. On OTel, the values are $J = 0.19$ versus $J = 0.36$. Across both SUTs, changing the prompt for a fixed model produces roughly half the Jaccard similarity of changing the model under a fixed prompt: prompt variation generates about twice the FM set diversity of model variation. The pattern holds for each model: \textcolor{black}{\texttt{qwen3:14b}} averages $J = 0.23$ (TeaStore) and $J = 0.15$ (OTel) across its prompt pairs, \textcolor{black}{\texttt{qwen2.5-coder:32b}} averages $J = 0.11$ and $J = 0.17$, and \textcolor{black}{\texttt{llama3.1:70b}} averages $J = 0.15$ and $J = 0.25$. All are below the cross-model same-prompt average on their respective SUT.

\begin{table}[t]
\centering
\caption{Average Pairwise Jaccard Similarity (Cross-Model, Same Strategy)}
\label{tab:jaccard}
\renewcommand{\arraystretch}{1.15}
\begin{tabular}{l c c}
\hline
\textbf{Strategy} & \textbf{TeaStore} & \textbf{OTel} \\
\hline
Structured     & 1.00 & 1.00 \\
CoT            & 0.30 & 0.49 \\
Guided         & 0.17 & 0.43 \\
FewShot        & 0.16 & 0.31 \\
GuidedFewShot  & 0.38 & 0.12 \\
Self-Refine    & 0.13 & 0.00 \\
\hline
\end{tabular}
\vspace{-12pt}
\end{table}

\subsection{RQ3: Effect of Domain Knowledge}
\label{sec:rq3}

\subsubsection{Rules Alone vs.\ Rules with Examples}

Guided prompting provides mutation rules without examples. GuidedFewShot adds the CRASH classification and examples of key-absent, value-empty, and value-null mutations. On both SUTs, GuidedFewShot achieves higher union coverage than Guided: 6 of 9 versus 4 of 9 on TeaStore (a gain of 2 FMs), and 9 of 14 versus 7 of 14 on OTel (also +2 FMs). The additional TeaStore FMs are FM$_3$ and FM$_6$, both key-absent mutations (\textcolor{black}{Finding~4}).

The pattern exposes a consistent limitation of LLMs. When the taxonomy says ``replace parameter value by null,'' all three models generate code that sets the parameter to an empty string. None omit the parameter key from the request body. These are different mutations exercising different code paths (Section~\ref{sec:ts_fm}), but LLMs treat them as equivalent unless shown examples of both. Both domain-knowledge strategies avoid the Structured collapse. Guided achieves $J = 0.17$ on TeaStore and $J = 0.43$ on OTel, compared to Structured's $J = 1.00$ on both (\textcolor{black}{Finding~5}). The mutation taxonomy provides direction without prescribing exact inputs, so each model applies the rules differently.

\subsubsection{\textcolor{black}{\texttt{qwen2.5-coder:32b}} Transformation}

The most dramatic effect of domain knowledge appears in \textcolor{black}{\texttt{qwen2.5-coder:32b}}, the code-specialized model. Under Self-Refine, \textcolor{black}{\texttt{qwen2.5-coder:32b}} triggers 0 FMs on both SUTs (\textcolor{black}{Finding~6}): the generated tests call endpoints correctly but omit the \texttt{assertNoServerError()} oracle, so every test passes trivially with no assertion checking the HTTP status code. \textcolor{black}{\texttt{qwen3:14b}} and \textcolor{black}{\texttt{llama3.1:70b}} include the oracle under Self-Refine and trigger 4 of 9 and 3 of 9 FMs on TeaStore, confirming that the oracle omission is specific to the code-specialized model rather than the Self-Refine strategy itself. GuidedFewShot achieves the study's highest single-run coverage: 5 of 9 FMs (56\%) on TeaStore and 8 of 14 FMs (57\%) on OTel (\textcolor{black}{Finding~7}). The swing from 0\% to 57\% is the largest performance change in the study.

\textcolor{black}{Code-specialized training gives \textcolor{black}{\texttt{qwen2.5-coder:32b}} the capability to produce syntactically correct HTTP test code, but robustness testing requires adversarial intent: inputs designed to break the system. That intent does not emerge from code training alone. The mutation taxonomy provides the adversarial framing, and the concrete examples show how to implement it in HTTP.}

\subsubsection{Unconstrained Generation Volume}

Guided and GuidedFewShot prompting do not specify how many tests to generate. Without this constraint, models produce varying volumes. On TeaStore under Guided, \textcolor{black}{\texttt{qwen2.5-coder:32b}} generates 25 tests versus 7 (\textcolor{black}{\texttt{qwen3:14b}}) and 9 (\textcolor{black}{\texttt{llama3.1:70b}}) (\textcolor{black}{Finding~8}). On OTel under GuidedFewShot, \textcolor{black}{\texttt{qwen2.5-coder:32b}} generates 60 tests, compared with 12 each for \textcolor{black}{\texttt{qwen3:14b}} and \textcolor{black}{\texttt{llama3.1:70b}}. On OTel under Guided specifically, \textcolor{black}{\texttt{llama3.1:70b}} generates 26 tests compared to \textcolor{black}{\texttt{qwen2.5-coder:32b}}'s 23, so the volume advantage is not universal across strategies. \textcolor{black}{\texttt{qwen2.5-coder:32b}} under GuidedFewShot on OTel produced over 2,000 lines of test code, entering a combinatorial loop over checkout credit card mutations; the output was truncated at line 683 for practical execution.

Higher volume does not automatically mean higher coverage. \textcolor{black}{\texttt{qwen2.5-coder:32b}}'s 60 OTel tests trigger 8 of 14 FMs (57\%), while \textcolor{black}{\texttt{llama3.1:70b}}'s 12 tests trigger 3 of 14 (21\%). The extra tests include redundant mutations of the same fields. However, higher volume increases the probability of encountering rare FMs: FM$_2$ (an extremely long product ID) appears only in high-volume runs.

\subsubsection{Coverage and Diversity Together}

On OTel, GuidedFewShot prompting achieves the highest single-run coverage (57\%), highest union coverage (64\%), and lowest non-zero Jaccard ($J = 0.12$). Each model finds a large and largely distinct FM set. Running all three provides substantial gains beyond any single run. Structured, by contrast, achieves 50\% single-run coverage on OTel with $J = 1.00$: additional models add nothing. On TeaStore, FewShot achieves higher union coverage than GuidedFewShot (89\% versus 67\%) because its three models collectively cover 8 of 9 FMs. GuidedFewShot's strength on TeaStore is single-run reliability: 56\% from one execution, with lower variance in what each model finds.

\begin{table*}[t]
\centering
\caption{Replicated Findings Across Both Systems Under Test}
\label{tab:findings}
\renewcommand{\arraystretch}{1.15}
\begin{tabular}{p{0.85cm} p{5.3cm} p{5.1cm} p{5.3cm}}
\hline
\textbf{ID} & \textbf{Finding} & \textbf{TeaStore Evidence} & \textbf{OTel Evidence} \\
\hline
\textcolor{black}{Finding~1} & No single run exceeds 57\% coverage & Max 56\% (5/9) & Max 57\% (8/14) \\
\textcolor{black}{Finding~2} & Prompt strategy $>$ model size for diversity & Confirmed & Confirmed \\
\textcolor{black}{Finding~3} & Structured collapse ($J = 1.00$) & All \textcolor{black}{models}: \{FM$_1$, FM$_8$\} & All \textcolor{black}{models}: same 7 FMs \\
\textcolor{black}{Finding~4} & Few-shot examples improve coverage over rules alone & Union: 6/9 vs.\ 4/9 & Union: 9/14 vs.\ 7/14 \\
\textcolor{black}{Finding~5} & Guided avoids Structured collapse & $J = 0.17$ vs.\ $1.00$ & $J = 0.43$ vs.\ $1.00$ \\
\textcolor{black}{Finding~6} & \textcolor{black}{\texttt{qwen2.5-coder:32b}} Self-Refine = 0 FMs (oracle omitted) & 0/9 FMs & 0/14 FMs \\
\textcolor{black}{Finding~7} & \textcolor{black}{\texttt{qwen2.5-coder:32b}}: 0\% $\rightarrow$ 56\%/57\% with domain knowledge & Self: 0/9 $\rightarrow$ GFS: 5/9 & Self: 0/14 $\rightarrow$ GFS: 8/14 \\
\textcolor{black}{Finding~8} & \textcolor{black}{\texttt{qwen2.5-coder:32b}} generates most tests unconstrained & 25 vs.\ \textcolor{black}{7 (\texttt{qwen3:14b}) and 9 (\texttt{llama3.1:70b})} & 60 vs.\ \textcolor{black}{12 (\texttt{qwen3:14b}) and 12 (\texttt{llama3.1:70b})} \\
\textcolor{black}{Finding~9} & Emergent security-aware inputs & \textcolor{black}{\texttt{qwen2.5-coder:32b}}-CoT: XSS payload & CoT: whitespace; \textcolor{black}{\texttt{llama3.1:70b}}-FS: XSS \\
\textcolor{black}{Finding~10} & GuidedFewShot = highest single-run coverage & 5/9 (56\%), \textcolor{black}{\texttt{qwen2.5-coder:32b}} & 8/14 (57\%), \textcolor{black}{\texttt{qwen2.5-coder:32b}} \\
\hline
\end{tabular}
\vspace{-12pt}
\end{table*}

\section{Discussion}
\label{sec:discussion}

\textcolor{black}{The results suggest different strategies based on available resources. Rather than prescribing a single configuration, we summarize practical guidance from our experience.}

\textit{Single model, single prompt:} GuidedFewShot is the strongest choice. It achieves the highest single-run coverage on both SUTs (56\% and 57\%) without requiring multiple models or multiple runs.

\textit{Single model, multiple prompts:} Varying the prompt strategy for one model is more effective than varying the model for one strategy. \textcolor{black}{\texttt{qwen2.5-coder:32b}} reaches 14 of 14 FMs (100\%) on OTel with just three strategies (Structured, GuidedFewShot, and ZeroShot). This is higher than any multi-model single-strategy union (maximum 64\% under GuidedFewShot). A practitioner with one model and limited time should run at least a domain-knowledge prompt (GuidedFewShot), a prescriptive prompt (Structured), and an unguided prompt (ZeroShot). \textcolor{black}{From a \textcolor{black}{reliability} perspective, the diversity of failure modes discovered by different configurations suggests that robustness testing with a single model-prompt pair provides an incomplete view of a system's failure surface. Multi-configuration testing is a practical step toward a more thorough assessment of the reliability of microservice deployments.}

\textit{Multi-model ensemble:} When multiple models are available, FewShot and GuidedFewShot provide the best return. FewShot reaches 89\% union on TeaStore; GuidedFewShot reaches 64\% union on OTel. Structured should be avoided when diversity is the goal: its $J = 1.00$ makes additional models redundant.

\subsection{Why Rules Need Examples}
\label{sec:rules_examples}

The mutation taxonomy from Vieira et al.~\cite{vieira2007} includes the rule ``replace parameter value by null.'' In HTTP, this maps to at least two distinct mutations: setting the value to an empty string (value-empty) and removing the key from the request body (key-absent). On TeaStore, the servlet's \texttt{request.getParameter()} returns an empty string for a value-empty parameter and \texttt{null} for a key-absent parameter, triggering different exceptions.

All three models, given only the taxonomy rules (Guided), produce only value-empty mutations. None produces key-absent mutations. This is consistent with how HTTP parameters appear in training data: form-encoded bodies almost always include parameter keys, and ``null'' in a programming context typically means assigning a null or empty value to a variable, not omitting the variable. The models' interpretation is reasonable given their training distribution, but wrong for robustness testing. GuidedFewShot resolves this by showing two concrete examples: an HTTP body with the key present and empty versus a body with the key absent entirely.

\textcolor{black}{The broader implication is that mutation rules written for human testers carry implicit assumptions about protocol-level implementation that LLMs do not share. Adapting domain knowledge for LLM prompts requires disambiguating rules that admit multiple valid implementations.}

\subsection{Exploratory Value of Unguided Generation}
\label{sec:exploratory}

A recurring pattern across both SUTs is that ZeroShot finds state-based FMs that no directed strategy discovers. FM$_9$ on TeaStore (unauthenticated session) and FM$_{14}$ on OTel (checkout without cart items) are both state-level failures: the system crashes not because of a bad parameter value but because of an invalid application state. Guided, GuidedFewShot, Structured, and FewShot all direct the model toward parameter mutations, which crowds out state-level thinking.

ZeroShot models, given only the API specification, generate tests resembling realistic user journeys: ``What happens if I remove an item without logging in?'' or ``What if I try to check out with nothing in my cart?'' These are the kinds of tests a manual tester might write from intuition rather than a mutation checklist. \textcolor{black}{The practical implication is that even a carefully designed testing pipeline should include at least one ZeroShot run. Its low per-run coverage (11\%--29\%) is offset by its ability to identify FMs missed by mutation-focused strategies.}

\subsection{Code-Specialized Models and Adversarial Intent}
\label{sec:model_b}

\textcolor{black}{\texttt{qwen2.5-coder:32b}}'s trajectory across strategies reveals a gap between capability and intent. It generates well-structured HTTP test code under every strategy. Under ZeroShot, it sends valid inputs. Under Self-Refine, it produces tests that call endpoints correctly but omits the oracle. Under GuidedFewShot, it achieves the study's highest FM coverage. \textcolor{black}{Code-specialized training optimizes for producing code that works, not code that deliberately breaks a system. The mutation taxonomy provides the adversarial framing that redirects this capability toward fault injection.}

The Self-Refine failure reinforces this point. During the critique step, \textcolor{black}{\texttt{qwen2.5-coder:32b}} evaluated its draft for code quality and endpoint coverage but did not flag the missing oracle assertion. Its self-evaluation criteria come from training on correct code, not testing methodology.

\subsection{The Structured Collapse}
\label{sec:collapse}

Structured achieves $J = 1.00$ because it serves as a test specification rather than a generation strategy. The prompt maps each endpoint to specific mutations and outcomes. All three models translate this table faithfully, producing functionally identical test suites. If a practitioner already knows which mutations to test, writing the tests manually or using a template-based tool is more efficient than prompting an LLM. The value of LLM-based test generation lies in the model's ability to produce mutations the practitioner did not anticipate. Structured prompting eliminates this value.

The contrast with GuidedFewShot is instructive. Both provide domain knowledge, but at different levels of abstraction. Structured prescribes specific inputs for specific endpoints. GuidedFewShot provides general mutation rules and a few examples, leaving the model to decide which rules apply to which endpoints. This higher abstraction preserves generative capacity while channeling it toward productive mutations.

\subsection{Emergent Security-Aware Generation}
\label{sec:security}

Several runs generated security-oriented inputs without instruction (\textcolor{black}{Finding~9}). \textcolor{black}{\texttt{qwen2.5-coder:32b}} under CoT on TeaStore generated XSS payloads (\texttt{<script>alert(1)</script>}) as product IDs. On OTel, CoT runs injected whitespace and control characters in checkout fields, and \textcolor{black}{\texttt{llama3.1:70b}}, under FewShot, generated XSS payloads for product IDs. None of the prompt strategies mentions security testing or injection attacks.

These inputs did not trigger robustness failures because the SUTs' input parsers rejected them before reaching application logic. \textcolor{black}{The observation suggests LLMs encode security testing patterns from training data, likely from exposure to OWASP documentation and penetration testing tutorials. This behavior is unpredictable and should not substitute for deliberate security testing, but it adds an unexpected dimension to LLM-generated robustness tests that warrants further study.}

\subsection{Beyond the Defined FM Set}
\label{sec:beyond}

\textcolor{black}{\texttt{llama3.1:70b}} under GuidedFewShot on OTel generated a cart quantity mutation using \texttt{Integer.MAX\_VALUE + 1}, triggering a failure outside FM$_1$ through FM$_{14}$. The mutation taxonomy includes boundary-value mutations for numeric types, and \textcolor{black}{\texttt{llama3.1:70b}} applied this rule to the cart quantity field, a parameter no other configuration targeted. \textcolor{black}{This suggests taxonomy-guided generation has value beyond reproducing known failures, potentially surfacing new failure modes. From a reliability perspective, this is encouraging: domain-informed LLM generation may help practitioners discover unanticipated failures in microservice deployments.}

\section{Threats to Validity}
\label{sec:threats}

\subsection{Internal Validity}
\label{sec:internal}

The primary internal threat is the verification of OTel checkout FMs. The currency and shipping services return HTTP 504 for all inputs in the available Docker deployment. Checkout calls both services internally, so HTTP 500 responses may reflect the 504 cascade rather than the input mutation. We classify checkout FMs (FM$_5$ through FM$_{14}$) by intended input mutation verified from source code. One checkout test under \textcolor{black}{\texttt{qwen2.5-coder:32b}} with GuidedFewShot returned HTTP 200, confirming the checkout service performs some input validation independently. TeaStore FMs are fully verified.

FM definitions are post hoc: derived from observed failures, not predefined. This risks overfitting the FM set to the models studied. \textcolor{black}{We mitigate this by requiring each FM to map to a specification-derived (endpoint, parameter, mutation-type) triple from the mutation taxonomy of Vieira et al.~\cite{vieira2007}, grounding the definitions in an established external reference rather than in the outputs of the models under study.}

Each configuration was executed once. LLM outputs are stochastic even under fixed decoding parameters~\cite{song2025}, and repeating the same prompt may yield different tests. We did not measure intra-configuration variance. \textcolor{black}{Findings based on a single run, such as \textcolor{black}{\texttt{qwen2.5-coder:32b}} omitting the oracle under Self-Refine, should therefore be interpreted as observations rather than reliable characterizations of model behavior. Multiple executions per configuration would allow variance measurement and strengthen confidence in individual findings, but were not feasible within the scope of this study. We recommend that future work treat single-run findings as hypotheses to be confirmed through repetition.}

\subsection{External Validity}
\label{sec:external}

Both SUTs are e-commerce applications \textcolor{black}{and open-source reference implementations rather than industrial systems}. Results may not generalize to other domains such as healthcare, financial services, or industrial control, where API structures and failure modes differ. \textcolor{black}{The two systems do, however, differ substantially in architecture, language stack, service count, and communication protocol, which supports the cross-system replication of findings reported in Table~\ref{tab:findings}.}

We evaluate three open-source models \textcolor{black}{with 14B, 32B, and 70B parameters}. Proprietary models (GPT-4, Claude, Gemini) and larger open-source models may produce different coverage and diversity patterns. \textcolor{black}{In particular, the finding that code-specialized training reduces adversarial intent may not hold for larger or recent code models.} Findings apply to the evaluated range and should not be extrapolated without further study.

The prompt strategies were designed by the study authors. Different formulations of the same strategy category (e.g., different few-shot examples or different CoT instructions) could change the results. \textcolor{black}{The Guided and GuidedFewShot strategies in particular embed specific choices about how to present the mutation taxonomy; alternative presentations may yield different coverage and diversity profiles.}

\subsection{Construct Validity}
\label{sec:construct}

The oracle classifies HTTP $\geq$ 500 as a robustness failure, corresponding to the Abort class in the CRASH taxonomy~\cite{vieira2007}. Silent failures (incorrect results with HTTP 200) and hindering failures (degraded performance) are not detected, so the true number of failure modes may exceed the 9 and 14 reported.

Jaccard similarity treats all FMs as equally important. A failure in checkout payment processing may have a greater practical impact than a failure in product ID lookup. \textcolor{black}{The current analysis does not weight FMs by severity; a weighted variant of the metric could reveal different diversity profiles and is left for future work.}

The binary coverage matrix records whether a run triggered a given FM at least once, not how many tests targeted it. A run triggering FM$_1$ once and a run triggering it ten times receive the same score. \textcolor{black}{This means that strategies producing high test volume, such as Guided and GuidedFewShot without a count constraint, are not rewarded for redundant coverage, which we consider a conservative and appropriate design choice.}

\section{Conclusion}
\label{sec:conclusion}

\textcolor{black}{We reported on applying LLM-based robustness test generation to two open-source microservice systems.} Through 38 experiments across 7 prompt strategies, 3 open-source models (14B--70B parameters), and 2 architecturally distinct microservice systems, we measured coverage over 23 failure modes and diversity via pairwise Jaccard similarity.

For RQ1, FM coverage varies substantially: no single run exceeds 57\%, but union coverage ranges from 11\% to 89\% depending on the strategy. A single model (\textcolor{black}{\texttt{qwen2.5-coder:32b}}) achieves 100\% FM coverage on OTel by varying the prompt strategy across just three configurations, outperforming any multi-model ensemble under a fixed prompt. For RQ2, diversity depends primarily on prompt strategy: Structured prompting collapses diversity entirely ($J = 1.00$), while FewShot and GuidedFewShot maintain substantially lower similarity, with GuidedFewShot ranging from $J = 0.12$ on OTel to $J = 0.38$ on TeaStore. For RQ3, mutation taxonomy rules from robustness testing research improve both coverage and diversity, but only when accompanied by concrete examples that disambiguate protocol-level mutations such as key-absent versus value-empty.

Several directions remain for future work. Evaluating proprietary and larger open-source models would test generalisability beyond the 14B--70B range. Applying the experimental design to non-e-commerce SUTs would assess domain generalisability. Running multiple executions per configuration would enable variance measurement and statistical significance testing. Resolving the OTel infrastructure limitation would enable full FM checkout verification. Automating FM definition from API specifications and source code would reduce the manual effort needed to establish ground truth.

\section*{Acknowledgment}
Claude (Anthropic) has been used for assisting editing. All experimental design, data collection, analysis, and interpretation were performed solely by the authors.

\bibliographystyle{IEEEtran}
\bibliography{references}

\end{document}